\documentclass{aa}
\usepackage{graphicx}\usepackage{rotating}
\newcommand{\etal}{{et~al.~}}

\newcommand{\integral}{{\it INTEGRAL }}

\newcommand{\rosat}{{\it ROSAT~}}

\newcommand{\xmm}{{\it XMM-Newton }}
\newcommand{\chandra}{{\it Chandra }} 

\newcommand{\asca}{{\it ASCA~}}

\newcommand{\kev}{\rm keV }
\newcommand{\flux}{\rm erg\ cm^{-2}\ s^{-1}}

\newcommand{\nh}{\rm cm$^{-2}$ }
\newcommand{\pks}{PKS~1830$-$211 }

\begin{document}

\authorrunning{A. De Rosa \etal.}
\titlerunning{The broad-band X-ray spectrum of \pks}

\title{The broad-band X-ray spectrum of the blazar PKS B1830$-$211 by 
\chandra and \integral}
\author{A.~De~Rosa\inst{1} 
\and L.~Piro\inst{1} \and A.~Tramacere\inst{2} \and E.~Massaro\inst{1,2} 
 \and R. Walter\inst{3} \and L.~Bassani\inst{4} \and A.~Malizia\inst{4} \and A.J. Bird\inst{5} \and A.J. Dean\inst{5} } 
\institute{
IASF - Sezione di Roma, INAF, via del Fosso del Cavaliere,
I-00113 Roma, Italy \and Dipartimento di Fisica, Universit\`a La
Sapienza, Piazzale A. Moro 2, I-00185 Roma, Italy \and Geneva Observatory, INTEGRAL Science Data Centre, Chemin d'Ecogia 16, 1291 Versoix, Switzerland \and
IASF - Sezione di Bologna, INAF, via P. Gobetti 101,
I-40129 Bologna, Italy \and School of Physics and Astronomy, University of Southampton, Highfield, Southampton, SO17 1BJ, UK }
\offprints{derosa@rm.iasf.cnr.it}
\date{Received ....; accepted ....}
\markboth{A. De Rosa \etal: The broad-band X-ray spectrum of the blazar 
PKS B1830$-$211 }
{A. De Rosa \etal.: The broad-band X-ray spectrum of the blazar PKS 
B1830$-$21}
\abstract{
In this paper we present a broad-band study of the X-ray emission of the
blazar \pks based on \chandra and \integral observations. 
Notwithstanding the high redshift ($z$=2.507), it is a bright X-ray 
source (F(2-10 keV)$\simeq 10^{-11}$ erg cm$^{-2}$s$^{-1}$), due to 
gravitational lensing by an intervening galaxy at $z$=0.89. 
Previous X-ray observations attribute the observed absorption at 
E$<$2 \kev to the lensing galaxy. Our analysis, although not in
contrast with this hypothesis, suggests also the possibility of 
an intrinsic (ionized) absorption, taking place at the front side of 
the jet. 
This scenario is also supported by some evidence, in the same data, of 
a feature observed at 2.15 \kev which can be interpreted as 
a blueshifted iron line ($v/c\simeq 0.18$). 
The SED of \pks can be well modelled by combining a Synchrotron 
Self-Compton component and an external source of photons 
to be scattered up to $\gamma$-ray energies by relativistic 
electrons moving outward in the jet. 
The main source of low energy photons is a dust torus at the
temperature of 10$^3$ K as expected in MeV blazars.
}
\maketitle


\section{Introduction}

PKS~B1830$-$211 is a high redshift blazar ($z = 2.507$, Lidman et al. 
1999) gravitationally lensed by an intervening galaxy at $z = 0.89$.
Its radio image shows two compact components separated by about 1$''$
 and believed to arise from the core of the source (hereafter North-East NE and
South-West SW components), plus a ringlike extended structure 
arising from the jet (Wiklind \& Combes 1996), connecting 
the compact components (Nair et al. 1993, Pramesh Rao \& Subrahmanyan 1988).
In the radio band the source shows strong variability (Lovell et al. 
1998) and has a flat spectrum. 
It was also observed in the infrared (Lidman et al. 1999), 
in the X-rays (Mathur \& Nair 1997, Oshima et al. 2001) and in 
the $\gamma$-rays (Mattox et al. 1997, Bassani et al. 2004).

$ROSAT$ data showed that the X-ray spectrum is best represented by a 
power law absorbed at the redshift of the lens galaxy by a column 
density $N_H=(3.5\pm0.5)\times 10^{22}$ cm$^{-2}$; no 
additional absorption component was required, suggesting that both lensed
images are covered by the same absorber (Mathur \& Nair 1997).
On the contrary, $ASCA$ data showed that the spectrum is consistent 
with two absorption components having different column densities:
$N_H^{low} < 1.5\times10^{22}$ cm$^{-2}$ and
$N_H^{high}$ = ($7.5\pm^{+0.8}_{-0.9}$)$\times10^{22}$ cm$^{-2}$
(Oshima et al. 2001), explained by the authors with the view that the 
low- and high-absorbing
component correspond to the NE and SW lensed images, respectively.
The ratio of the two magnification factors was
0.21$^{+0.07}_{-0.08}$, smaller than the value found at 
radio wavelengths which was 1.55$\pm0.02$ (Lovell \etal 1998).
In this paper we investigate the broad-band X-ray spectrum of \pks
observed by \chandra and \integral.
In section \ref{data} we describe the observations while data analysis 
is presented in Section \ref{analysis}. In
section \ref{absorber} the nature of the complex absorber is
discussed, while in Section \ref{sed} the SED of \pks is derived
and a possible emission model is presented.
We summarize our results and conclusions in Section \ref{conclusions}.

\begin{figure*}[!t]
\centering
\includegraphics[height=5.5cm,width=16.5cm]{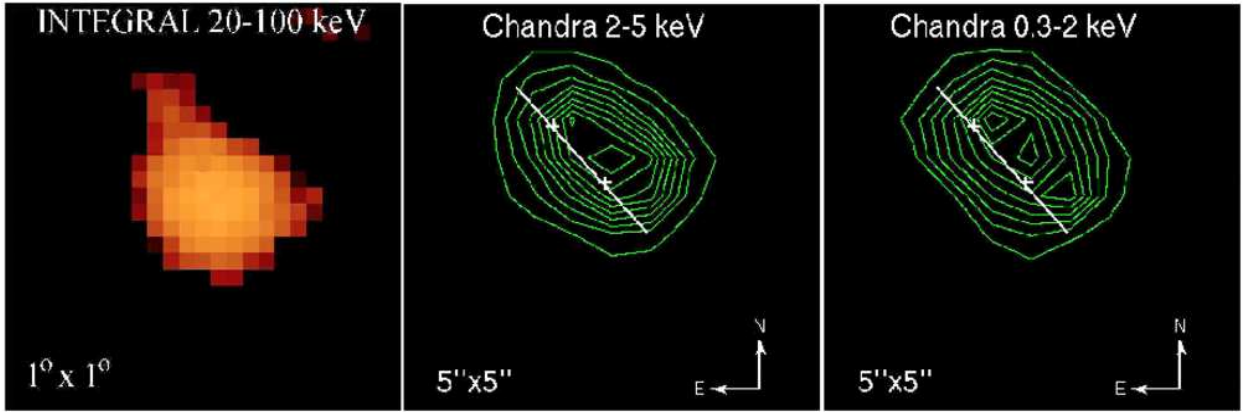}
\caption{X-ray images of \pks. The \integral image is shown in
    the left panel, its extension is compatible with the PSF of ISGRI. 
    In the middle and right panel we show the \chandra contour levels in two
    different energy range together with the position of the NE and SW
    lobes from radio observations (Pramesh Rao \& Subrahmanyan
    1988). The two main components of the lensing are distinguished in 
    \chandra images.}
\label{images}
\end{figure*}

\section{Observations and data reduction}
\label{data}

\begin{figure}[!b]
\centering
\includegraphics[height=8.5cm,width=9.0cm]{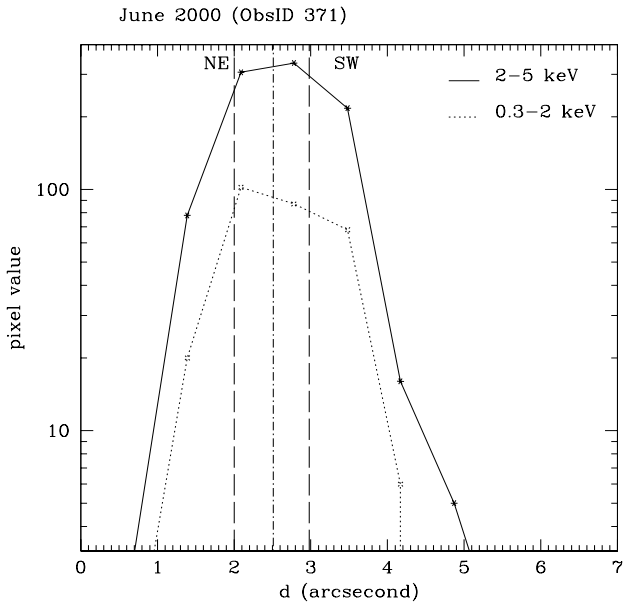}
\caption{X-ray intensity profiles along the
major axis of the ellipse in the case of the \chandra observation in 2000 June.}
\label{profile}
\end{figure}

\subsection{INTEGRAL}
PKS1830-211 is one of the AGNs detected by $INTEGRAL$ in the Galactic Center
region (Bassani et al. 2004). It has been reported as an $ISGRI$ source 
by Bird et al. (2004) and Revinivtsev et al. (2004) and at a redshift of 
2.507 it is the farthest object so far detected by $INTEGRAL$.
The observational data reported here refer to the first $IBIS$ survey which 
consists of several pointings carried out between 2003 February 28 to October 
10. 
The total on source exposure is 714 ksec, providing a detection at 
$\sim$14$\sigma$ confidence level. 
The mean counting rate is 0.29${\pm}$0.03 counts s$^{-1}$ both in the 20--40 
and 40--100 \kev band, a clear indication of a hard spectrum. 
These count rates correspond to a flux of roughly 3 (2.4) and 4 (3.9) mCrab
(10$^{-11}$ erg cm$^{-2}$ s$^{-1}$) in the two bands, respectively.
 ISGRI images for each available pointing were generated in  narrow energy
bands using the ISDC offline scientific analysis software OSA version
3.0 (Goldwurm \etal 2003), including
background uniformity corrections (Terrier \etal 2003). Source
ghosts have been removed from each image using a catalogue built iteratively
and containing at end all sources detected. The individual images were then
combined to produce a mosaic of the region of interest in broader bands to
enhance the detection significance using the system described in details by
Bird \etal (2004).

The combined 20-100 keV image of the source is shown in the left panel in 
Figure \ref{images}.
The ISGRI image is point-like, the extension is compatible with
the PSF of the instrument.

\subsection{Chandra}
$Chandra$ observations of PKS~B1830$-$211 were performed on two 
occasions, one year apart in 2000 June 26-27 and 2001 June 25, 
for a total exposure of about 50 ks. 
The source did not show evidence of variability between the 
two pointings. 
The good spatial resolution allow us to observe the two lensed 
images, with an angular distance of the order of the \chandra 
resolution limit ($\sim1''$).
In Figure \ref{images} we show the ACIS contour levels extracted in two 
separate energy ranges 0.3-2 keV (right panel) and 2-5 keV (middle 
panel). It the figure we indicate also the position of the NE and SW
component taken from radio observation (Pramesh Rao \& Subrahmanyan
1988). The offset between radio and X-rays image is less than 1$''$. 
There is evidence for an ellipsed shape in both images, as showed 
by the plotted contours, and the two main components of the lensing
effect are visible. 
In Figure \ref{profile} we plot the X-ray intensity profiles along the
major axis of the ellipse which shows that it is significantly
larger than the ACIS PSF. The ACIS PSF is very sharply peaked
on-axis, with a FWHM less than half an arcsec\footnote{See {\it The Chandra
  Proposers Observatory Guide 
(http://asc.harvard.edu/proposer/POG/html/MPOG.html)}}.
The ratio between the counts from the NE and SW 
components is about 1.5 in both 0.3--2 keV and 2--5 keV energy range
similar to the value observed at radio frequencies.
This result indicates that the absorbing gas column densities of 
the two components cannot be very different, contrary 
the findings of Courbin et al. (2002) in the optical and near-IR, 
where the SW component seems to be much weaker than the NE one.

\section{Spectral analysis and results}
\label{analysis}

\begin{figure}[t]
\centering
\includegraphics[height=8.5cm,width=9.0cm]{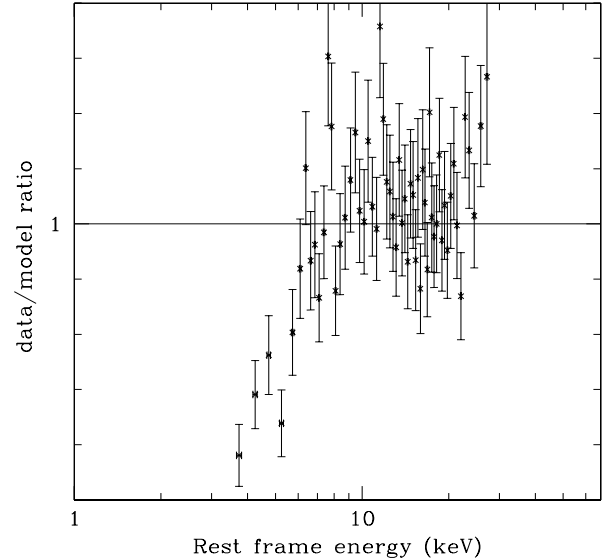}
\caption{Ratio of \chandra data of both observations to 
a $\Gamma=1.02$ simple power-law (including Galactic 
absorption $N_H^{Gal} = 2.6\times10^{21}$ \nh and fitted over the
2--8 \kev~ range) in the source frame.
Note the large residuals at energies below 5 \kev.
}
\label{ratio}
\end{figure}

Spectral analysis has been performed fitting the data of both \chandra 
observations combined together, in fact fitting them separately does not 
provide evidence for a significant difference.
We first apply a simple power--law absorbed by Galactic gas 
($N_H^{Gal} = 2.6\times10^{21}$ \nh, \cite{stark92}) in the whole range 
0.5--8 \kev and obtain a poor fit with $\chi^2$/dof=725/459.
A better result is found fitting the data at energies higher than 
2 \kev, where the absorption is not relevant. 
In this case the above model gives a $\chi^2$/dof=389/377 with 
a photon index $\Gamma$=1.02$\pm$0.05 and an estimated flux  
$F_{2-10} \simeq~10^{-11} \flux$.
However, the data to model ratio in the energy range 
2--8 \kev shows clear systematic residuals below 5 \kev
in the source frame (Figure \ref{ratio}).
\integral spectrum in the hard X-ray energy range 20-80 \kev is also 
well reproduced by this flat power--law model,
confirming that the broad--band 
spectrum remains remarkably flat up to 80 keV. This will be discussed in 
more detail in the next Section.

\section{The nature of the absorption}
\label{absorber}
Spectral flattening at energies less then 2 \kev~ has been 
found in radio-loud quasars having a redshift up to 4.4, using \asca and
\xmm data (Fiore \etal 1998, Reeves \& Turner 2000, Worsley \etal
2004, Yuan \etal 2003 and references therein). 
Absorption has been suggested as the likely explanation for this flattening but
the data collected up to now do not allow one to distinguish between
an intrinsic absorber at the source redshift or an absorbing
system along the line of sight.
However, the UV-optical extinction observed in most of these objects,
is inconsistent with the high column density required to produce the
flattening observed in X-rays. It was suggested that this gas
could be ionized (Fabian \etal 2001), i.e. a {\it warm absorber}
similar to that  present in Seyfert 1 galaxies.
Another possibility is that of an intrinsic low energy cut-off 
in the electron spectrum as proposed for other similar objects
(Celotti \& Fabian 1993, Sikora \etal 1997).

\begin{table*}[t]
\caption[] {Results of the spectral fits of \chandra and \integral observations. Confidence ranges are 90 per cent for one parameter.}
\begin{flushleft}
\begin{tabular}{l l l l l l l l}
& & & & & & & \\
\hline
 & $\Gamma$ & $^1N_{H,z}$/$^1N_{H,ion}$ & $^2\xi$ & $^{3}F^{una}_{2-8 keV}$ & $^{3}F^{una}_{2-80 keV}$ &  $^{3}F^{una}_{0.5-8 keV}$ & $\chi^2$/dof \\
& & & & & & & \\ \hline
0.5-8\kev & $1.09^{+0.05}_{-0.05}$ & $1.93^{+0.27}_{-0.26}$ & - & 0.78 & 8.6 & 1 & 482/458\\
& & & & & & & \\
0.5-8\kev &$1.06^{+0.05}_{-0.05}$ &$9.5^{+4.4}_{-2.1}$ & $<100$ &0.78 & 9.1 & 1 & 517/457\\
& & & & & & & \\
0.5-80\kev & $1.09^{+0.05}_{-0.05}$ &$1.94^{+0.28}_{-0.25}$ &-&0.78 &8.5 &1 & 485/462\\
\noalign {\hrule}
\noalign{\medskip}
\noalign{\noindent
Note: $^{(1)}$ $N_{H,z}$ is the column density at redshift of the lens galaxy 
$z$=0.89, $N_{H,ion}$ in the column density of the warm gas at redshift of 
the source $z$=2.51, both are in $10^{22}$ cm$^{-2}$;  
$^{(2)}$ in {\rm erg cm s$^{-1}$};
$^{(3)}$ in {$10^{-11}$ $\rm erg\ cm^{-2}\ s^{-1}$ }}
\end{tabular}
\end{flushleft}
\label{fit}
\end{table*}

\begin{figure*}[t]
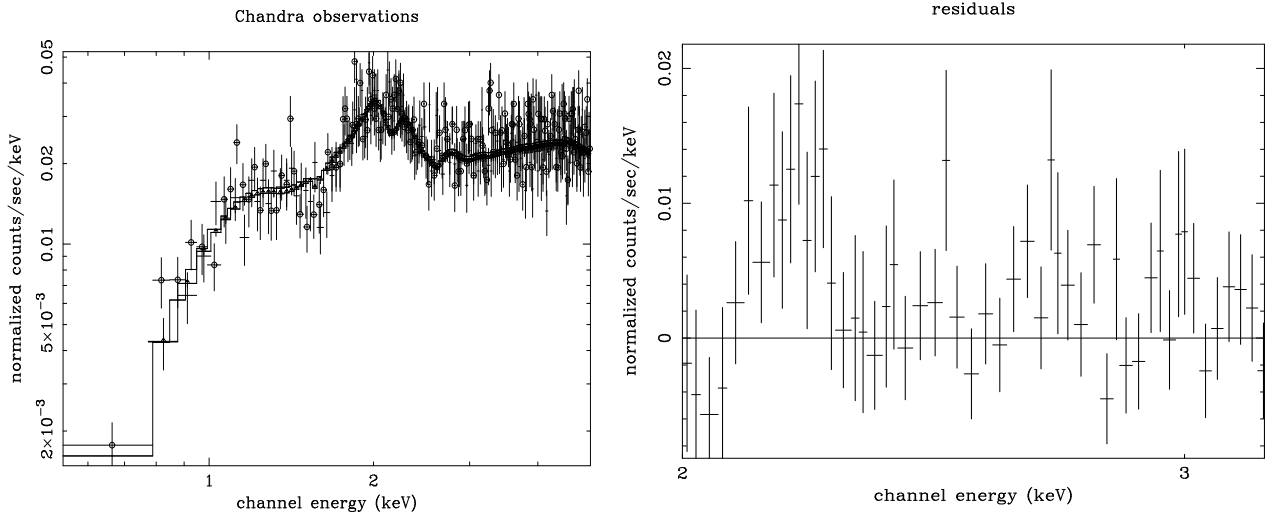

\includegraphics[height=7.5cm,width=7.cm,angle=0]{pk-fig4a.ps}
\includegraphics[height=7.5cm,width=7.cm]{pk-fig4b.ps}
\caption{In the left panel we show the \chandra spectra of both 2000 
and 2001 observations with the folded model characterized by a 
power--law absorbed by a cold gas at the redshift of the lens galaxy. 
The data taken in 2001 are plotted with a circle.
A zoom of the residuals in the line feature region, from 
\chandra ACIS-S, is shown in the right panel.}
\label{chandra spec}
\end{figure*}

\begin{figure}[t]
\centering
\includegraphics[height=8.5cm,width=8.0cm]{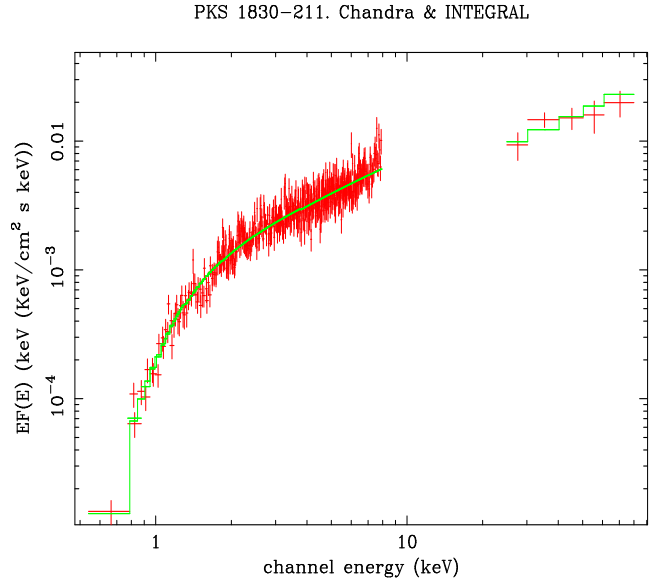}
\caption{Broad--band spectra of \pks from \chandra
  and {\it INTEGRAL}. The observations have been performed about one year
  apart. The data are reproduced with a power-law with $\Gamma\simeq$ 
  1.1 absorbed at $z$=0.89 by $N_H\simeq 2\times10^{22}$ \nh (in 
  addition to the Galactic absorption at $z$=0).}
\label{bbspec}
\end{figure}

First, we attempted to reproduce the low-energy spectral break 
with a cold absorber (in addition to the Galactic one) at the redshift of 
the lens galaxy ($z$=0.89) 
as proposed by  previous \rosat (\cite{mathur97}) and 
\asca (\cite{oshima01}) observations. 
The best fit values are reported in the first line of Table \ref{fit}.
\chandra data of both observations are well fitted with a single 
flat power-law, 
with photon index $\Gamma$=1.09$\pm$0.05,
absorbed by a cold gas with N$_H$ $\simeq$2$\times$10$^{22}$ \nh 
at $z$=0.89 ($\chi^2$/dof=482/458). 
Data and the folded model are shown in Figure \ref{chandra spec} (left panel).
\integral data, taken one year after \chandra observation, are also 
well reproduced by this continuum. 
The constant of cross--calibration \chandra/\integral was free to vary 
during the fit and was found to be C=0.57$\pm$0.13. 
The broad--band spectrum of 
\pks is shown in Figure \ref{bbspec}, and the best fit values are reported 
in the last line of Table \ref{fit}.\newline
We also test the possibility that the absorbing gas is intrinsic to 
the source, by fitting \chandra data with a warm absorber model. 
Best fit values are reported in the second line of Table \ref{fit}. 
The fit is good ($\chi^2$/dof=517/457) even if 
not better with respect to that of a cold gas at $z$=0.89. 
This result suggests the presence of a mildly ionized gas with
the column density $N_H\simeq10^{23}$ \nh\,,
and an upper limit to the ionization parameter $\xi=L_{2-10}/nR^{2}<100$ 
{\rm erg cm s$^{-1}$}, where $L_{2-10}$ is the inferred isotropic 
luminosity in the 2-10 \kev~ rest frame of the blazar while $n$ and $R$ 
are the absorber density and the distance from the source,
respectively.

\chandra data show some evidence ($\approx 3 \sigma$) in each 
observations of an emission feature around 2.15 \kev~ with an
intensity of $2\times10^{-5}$ {\rm ph cm$^{-2}$ s$^{-1}$} and an
equivalent width of about 50 {\rm eV} (see right panel in Figure 
\ref{chandra spec}). 
The line is not resolved in the ACIS-S, setting the intrinsic 
width to be $<50$ eV. 
\chandra-HETG spectra are consistent with ACIS data, and set a
lower limit on the intrinsic width to be 10 eV.  
It is important to note that the residuals around the line are of the 
order of few tens of per cent, while the instrumental systematic effects at 
this energy are of the order of few per cent.
The energy of the line at the redshift of \pks is 7.54 \kev~
thus suggesting an association with a blueshifted iron line. 
This feature could be produced in the ambient matter 
around the source.
If the line is due to neutral iron, the implied velocity for the 
gas has to be $v\simeq0.18c$.\
It is interesting to note that if the warm gas with a column
density of $\sim$ 10$^{23}$ \nh covers a solid angle of 2$\pi$ to
the source, it should produce a line with EW $\simeq$ 50 eV,
consistent with that detected in \chandra spectrum.

\section{The Spectral Energy Distribution}
\label{sed}
\begin{figure*}
\centering
\includegraphics[height=11.0cm,angle=-90]{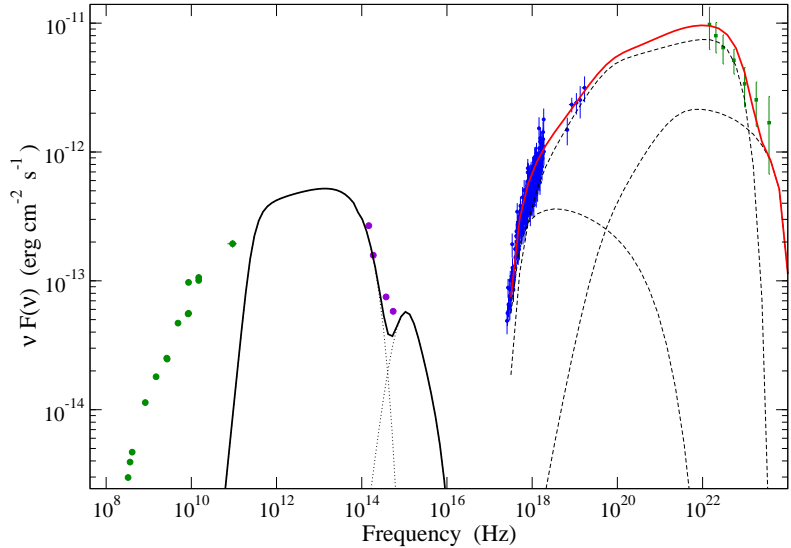}
\caption{
The Spectral Energy Distributions of \pks in the observers' frame 
derived from our results and other published data. 
Radio measurements are from Pramesh Rao \&
Subrahmanyan (1988), IR and optical data from Courbin et al. (2002) 
and $\gamma$-ray data from the EGRET public data archive.
Solid lines represent the SED as the sum of various emission 
contributions computed with the parameters given in the text;
dotted lines are the individual low energy contributions: 
synchrotron emission and accretion disk; dashed line are 
high energy spectra of SSC (lower bump) and ERC components.
}
\label{sed fig}
\end{figure*}

We then used our spectral results and other data available in the 
literature to derive the SED of \pks shown in Figure \ref{sed fig}.
Of course, it is not based on simultaneous observations and
therefore must be considered an indicative picture of the emission
properties of this source. 
$\gamma$-ray data are from the EGRET public archive 
(ftp://cossc.gsfc.nasa.gov/compton/data/egret/) and indicate
a quite steep spectrum with a photon index of 2.58$\pm$0.13. 
IR and optical points have been derived from the photometric
data of the NE component given by Courbin et al. (2002), multiplied 
by 2 to take into account the obscured contribution from the SW 
component.
These data have been corrected for an equivalent galactic absorption 
$A_V$=6.0, slightly smaller than the $N_H$ value expected from the 
\chandra data best fit. The galactic contribution to $A_V$, estimated from 
$E(B-V)$=0.46 (Schlegel et al. 1998), is 1.44 ($R=3.1$), whereas 
the additional absorption from the lens galaxy is of the order 
of 4 mag (Winn et al. 2002).
We recall that optical images show that the SW component is much
higher absorbed with a differential reddening of $E(B-V)\simeq 2.75$
(Courbin et al. 2002).

A further problem is the magnification factor of the flux due to
the gravitational lens. Following the numerical calculations for
the lens modelling by Nair \etal (1993) we assumed the magnification
factor for the summed flux of both components to be equal to 10 (see also 
Mathur \& Nair 1997). 
A lower factor, of course, would increase the intrinsic luminosity 
of PKS~1830$-$211.
 
We then tried to apply some emission models to reproduce the SED 
according to the largely accepted scenarios used for high luminosity 
Blazars.
First, we used a single zone homogeneous Synchrotron-Self 
Compton (SSC) model, but it failed to reproduce at the same 
time the radio-optical bump and the X and $\gamma$-ray bump. 
The SSC model contains a number of free parameters whose values were
chosen on the basis of the current blazar physics: we assumed then
the beaming factor $\delta$=16 and a mean magnetic field $B$=0.8 G,
comparable to those considered for 3C~279 (Hartman et al. 2001).
The very high beamed luminosity, $\sim$4$\times$10$^{48}$
erg s$^{-1}$ after the correction for the lens magnification, 
requires a population of emitting electrons confined inside a volume 
of about $10^{53}$ cm$^3$ and an electron density of 850 cm$^{-3}$.
The steady electron energy distribution is a broken power law with the break 
at the Lorentz factor $\gamma_b \simeq$ 70, obtained equating 
the electron cooling time to the crossing time through the 
emitting volume.
The two spectral indices are $p_1=1.80$ for $\gamma<\gamma_b$ and
$p_2=2.80$ above the break. These values were chosen to match together
X and $\gamma$-ray data are their difference is that expected from 
radiative cooling.  
The maximum Lorentz factor is 2500, suited to reproduce the decline of
synchrotron spectrum at optical-UV.

The resulting SED is plotted in Figure \ref{sed fig}. Note that the synchrotron
self absorption frequency (defined by unitary optical depth) is 
around 3$\times$10$^{11}$ Hz in the observer's frame and therefore 
there is no way to account for the emission in the GHz band. 
The same problem, found for other blazars like 3C 279 (Hartman et 
al. 2001) and PKS~1127$-$145 (Blazejowski et al. 2004), can be solved 
assuming a multicomponent model in which the radio emission
comes from an outer region where the magnetic field is lower.
Notice, however, that this SSC model gives hard X and $\gamma$-ray 
luminosities lower than the observed values by more than an order 
of magnitude.

External Radiation Compton (ERC) models consider that the main sources 
of seed photons are the accretion disk around the massive black hole,
the broad line region and, at larger distances, the dust torus. 
UV photons from the disk are seen deboosted by the relativistic electrons
moving outward in the jet from behind and their contribution is small.
When computing ERC considering only photons originating in an
accretion disk with a luminosity of 4$\times 10^{48}$ erg s$^{-1}$ 
and a multitemperature spectrum ($T_{max}=3.5 \times 10^5$ K, Shakura 
and Sunyaev 1973) we obtain a Compton bump much lower than the data.
Moreover, photons scattered by the BLR give another relevant but not
a sufficient contribution to IC bump. An increase of $\gamma_b$ would
move the peak frequency above 10$^{23}$ Hz which is in conflict with
the steep spectrum observed in the EGRET range.

More suitable are photons produced by matter at a distance of a few
pc and illuminated by the disk, like the IR emission from a dusty torus
at a temperature of 10$^3$ K as expected in similar sources (Sikora et 
al. 2002, see next section).
Given the geometry of the jet-disk ambient, and the positioning of the 
emitting volume at a distance of $\sim$10$^{19}$cm from the accretion 
disk, we find that the comoving energy density of the IR external 
photon field from the dust torus overwhelms the other soft field at 
least by an order of magnitude.
The resulting IC bump is able to reproduce both the X and $\gamma$-ray
continuum. We recall, however, that the latter data are not a strong
constraint because of source variability, which in this range is 
at least a factor of 4, as indicated from the 3EG catalogue 
(Hartman et al. 1999).

\section{Discussion}
\label{conclusions}
It is interesting to compare the properties of this distant and
lensed blazar with those of other sources of the same class.
The SED of Figure \ref{sed fig} shows that the largest energy output is in the
prominent IC bump.
A relevant finding of our analysis is that the X-ray spectrum 
of \pks is remarkably flat up to energies of $\sim$80 keV,
while the spectrum above $\sim$100 MeV is much steeper.
The difference between these two spectral indices is around 1.5.  
The peak frequency, however, is not well constrained by the data: in 
the model described in the previous section it lies around 10$^{22}$ 
Hz, however it could be lower, in the MeV range, for a different 
electron spectrum. 
\pks can be considered a member of the small subclass of ``MeV 
blazars'' discovered by COMPTEL (Bloemen et al. 1995), which are 
characterised by flat hard X-ray and rather steep $\gamma$-ray 
spectra (Blom et al. 1996, Tavecchio et al. 2000).
Sikora et al. (2002) proposed that the prominent IC bump seen in 
this type of blazars can be explained by ERC dominated by the 
interactions of relativistic electrons with the near-IR photons 
from a hot dusty torus.
Our SED modelling of the previous section confirms this scenario
also in the case of PKS~1830$-$211 
SSC emission is indeed found not enough efficient to reproduce 
the IC bump and the ERC emission on optical-UV photons from the 
BLR gives too much energetic $\gamma$ rays.

A complete understanding of the emission properties of PKS~1830$-$211, 
however, presents some difficulties arising from the presence of 
the gravitational lens. Radio, optical and X-ray data do not give
a consistent picture on the absorption of the two lensed components.
The large brightness difference observed by Courbin et al. (2002)
in $HST$ images was explained by a differential extinction 
$\Delta A_V \simeq$~8.5. One can expect that X-ray images at
energies lower than 2 keV would show a SW component much weaker 
than the NE, however this effect is not observed in our data 
(see Section \ref{data}). 
Spatially resolved high resolution radio spectroscopy of HCO and 
HCN transitions at the lens redshift (Swift et al. 2001) showed 
that there is no absorption in the NE image, while the SW one
presents optically thick absorption with a complex structure. 
Molecular clouds on the radiation path of the SW component can
be then responsible of the differential absorption, however the 
dust to gas density ratio should be lower than the mean galactic 
value to account for the much higher optical extinction. 

The spectral behaviour of \pks at low X-ray energies is characterized by 
strong absorption.
In Figure \ref{ratio} we show the spectrum, 
in the rest frame, of the combined \chandra observations as the ratio 
data to model assuming a power--law absorbed by Galactic gas. 
The ratio is not effected by instrumental response and Galactic absorption.
We found that the spectrum below 5-6 \kev (source frame) flattens 
indicating absorption in excess of the Galactic one, probably due to the 
lens galaxy at $z$=0.89 and with a column density N$_H\sim$10$^{22}$ \nh.
However a remarkably similar shape is shown in Figure 5 of Yuan \etal 
(2003) in the case of the \xmm observations of  
RX J1028.6$-$0844 ($z$=4.276), GB1428+4217 ($z$=4.72) and PMN J0525$-$3343 
($z$=4.4).
The energy of the break is similar in all sources, with a 
power-law photon index $\Gamma\simeq 1$.
This evidence strongly encourage a common scenario in all these high redshift 
sources where absorption plays a role.
The most plausible hypothesis (Fabian \etal 2001) includes the presence 
of a warm gas in the nuclear region 
(similar to that observed in Sy 1 galaxies).
This warm gas is intrinsic to the source and is characterized by a 
column density N$_{H,z}\sim$10$^{23}$ \nh.
In the case of \pks this model seems to be supported by a marginal evidence 
of an emission line at the observed energy 2.15 \kev, that can be interpreted 
as a blueshifted iron line ($v/c\simeq 0.18$).

Further \chandra and \xmm observations will help to understand the 
nature of the absorption in \pks and in the other radio--loud quasars 
at high redshift.

\begin{acknowledgements}

ADR would like to thank A. Bazzano and P. Ubertini for useful discussions.
We acknowledge financial support by ASI (Italian Space Agency) via contract
I/R/041/02. Part of this work was performed with the financial 
support Universit\'a La Sapienza di Roma.

\end{acknowledgements}

\end{document}